\documentstyle [11pt,paspconf]{article}
\input{epsf.tex}
\begin{document}
\par\noindent \hfill
 an invited talk at: {\it The International Conference on Stellar
Dynamics: \par\hskip .35in \hfill
From Classical to Modern}, Sobolev Astronomical Institute, 
\par\hskip .35in \hfill
St. Petersburg State University, August 2000
\vskip .2in
\title{Chaotic Mixing in Galactic Dynamics}
\author{Henry E. Kandrup}
\affil{Department of Astronomy, Department of Physics,
and Institute for Fundamental Theory,
University of Florida, Gainesville, FL 32611}

\begin{abstract}
This talk summarises what is currently understood about the phenomenon that 
has come to be known as {\it chaotic mixing}. The first part presents a 
concise statement as to what chaotic mixing actually is, and then explains why 
it should be important in galactic dynamics. The second discusses in detail
what is currently known, describing the manifestations of chaotic mixing 
both for flows in time-independent Hamiltonian systems and for systems 
subjected to comparatively weak time-dependent perturbations. The third part 
discusses what one might expect if one were to allow for a strongly 
time-dependent potential, including possible implications for violent 
relaxation.
\end{abstract}
\vskip .3in
\section{What Is Chaotic Mixing and Why Does it Matter?}
Initially localised ensembles of initial conditions corresponding to chaotic
orbits will, when integrated into the future, disperse exponentially at a rate
set by the largest (short time) Lyapunov exponents ${\chi}$. By contrast, 
ensembles of initial conditions corresponding to regular orbits will only 
disperse as a 
power law in time. This much is obvious: After all, the defining characteristic
of chaotic orbits is that they exhibit exponentially sensitive dependence on
initial conditions. Not so obvious, perhaps, but also true is the following:
At least in time-independent potentials, initially localised ensembles of 
chaotic initial conditions tend to evolve exponentially in time towards an 
invariant or near-invariant distribution, {\it i.e.,} an equilibrium or
near-equilibrium ({\it cf.} Kandrup \& Mahon 1994, Mahon {\it et al} 1995,
Kandrup 1998). In particular, as probed by lower order moments or by 
coarse-grained distribution functions, an initially localised ensemble of 
chaotic orbits typically evolves exponentially in time towards a near-uniform
population of those phase space regions that are easily accessible, {\it i.e.,}
not impeded by cantori or an Arnold web.

This means that phase mixing can proceed {\it much} more efficiently 
for chaotic flows than for regular flows, where any approach towards a 
(near-) equilibrium typically proceeds as a power law in time.
Chaotic flows should relax much more efficiently than do regular flows. It
would thus seem that the phase mixing of chaotic flows, a phenomenon which 
Merritt \& Valluri (1996) have termed {\it chaotic mixing}, could serve to
provide an explanation of why various systems in nature seem to approach an
equilibrium or near-equilibrium as fast they do. In particular, chaotic mixing
could help explain the remarkable efficacy of violent relaxation: Why do 
galaxies look `so relaxed' when the nominal relaxation time $t_{R}$ is 
typically much longer than $t_{H}$, the age of the Universe? And, in 
particular, why can galaxies recover from a collision or close encounter as
quickly and as efficiently as they apparently do?
\section{Manifestations of Chaotic Mixing}
\subsection{Time-independent Hamiltonian systems}
A localised ensemble of initial conditions corresponding to chaotic orbits
will, when evolved into the future, begin by diverging in such a fashion that
quantities like the dispersions in position or velocity diverge exponentially
at a rate set by a characteristic short time Lyapunov exponent for the orbits,
so that, {\it e.g.,} ${\sigma}_{x}(t){\;}{\propto}{\;}\exp(+{\chi}t)$. 
Eventually, however, this divergence saturates, so that such quantities
asymptote towards a near-constant value, which would suggest that the ensemble 
is approaching some equilibrium, or near-equilibrium, state. That this is 
actually so can be demonstrated by tracking the evolution of coarse-grained 
distribution functions. Thus, {\it e.g.,} orbital data at fixed instants of 
time can be binned onto a $k\times k$ grid so as to construct coarse-grained 
distribution functions like $f(x,y,t)$ or $f(x,v_{x},t)$, and one can then ask 
whether it be true that such $f(t)$'s converge towards nearly time-independent 
distributions $f_{niv}$. To address this question, one requires a definition 
of `distance' between two distributions $f_{1}$ and $f_{2}$. The obvious 
choice entails a `pixel by pixel' comparison in terms of a discrete $L^{p}$ 
norm, so that, {\it e.g.,} the distance between $f_{niv}(x,y)$ and some 
$f(x,y,t)$ satisfies
\begin{equation}
Df(x,y,t)=\left[
{\sum_a\sum_b \left| f(x,y,t) - f_{niv}(x,y) \right|^{p} \over
 \sum_a\sum_b \left| f_{niv}(x,y) \right|^{p}} \right]^{1/p}
\end{equation}
with $p=1$ or $2$.

Numerical experiments (Kandrup \& Mahon 1994, Mahon {\it et al} 1995,
Merritt \& Valluri 1996, Kandrup 1998) indicate that, 
with respect to this measure of distance,
$f(t)$ does indeed converge towards an invariant, or near-invariant, $f_{niv}$;
and that, at least early on, this convergence is well fit by an exponential, 
{\it i.e.,}
\begin{equation}
Df(t){\;}{\propto}{\;}\exp(-{\Lambda}t) .
\end{equation}
This is illustrated in Figure 1, which exhibits ${\sigma}_{x}(t)$
and $Df(x,y,t)$ for two different ensembles evolved in the three-dimensional
dihedral potential ({\it cf.} Kandrup 1998). The `saturation' observed in the
plots of $Df(t)$ is (at least primarily) a finite $N$ effect: even if
the coarse-grained $f(t)$ and $f_{niv}$ involved two different $N$-orbit 
samplings of the same smooth distribution, they would necessarily differ
because of finite number statistics.
\begin{figure}[t]
\centering
\centerline{
        \epsfxsize=8cm
        \epsffile{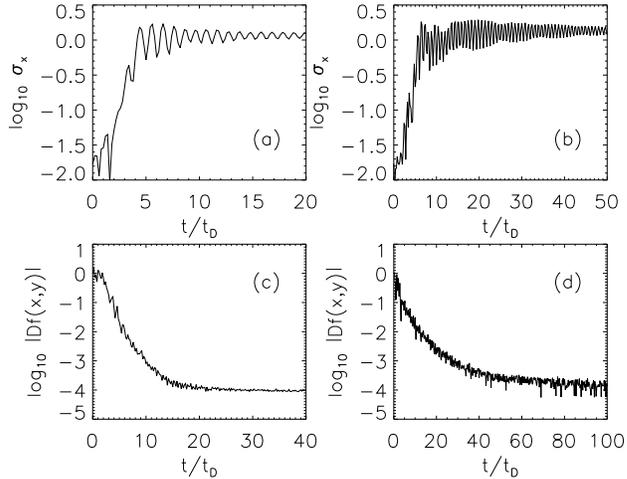}
           }
        \begin{minipage}{12cm}
        \end{minipage}
        \vskip -0.4in\hskip -0.0in
\caption{ %FIG. 1 
\small
${\sigma}_{x}(t)$ and $Df(x,y,t)$ for two 
different ensembles of chaotic orbits, with $E=1$ and $E=4$, evolved in the
three-dimensional dihedral potential with $a=b=1$.} %\label{fig-1}
\vspace{-0.2cm}
\end{figure}
Because the rate at which moments like ${\sigma}_{x}$ grow is set by the 
Lyapunov exponents, one might speculate that ${\Lambda}$ should also be related
to ${\chi}$. However, there does {\it not} always seem to be a clean 
one-to-one connection. One {\it does} find numerically that larger ${\chi}$ 
correlates with larger ${\Lambda}$, and that ${\chi}$ is always larger than 
${\Lambda}$, {\it i.e.,} the rate at which orbits diverge is larger than the 
rate at which they approach an equilibrium. However, there does not appear to 
exist a universal connection between ${\chi}$ and ${\Lambda}$ that holds for 
generic potentials admitting a coexistence of both regular and chaotic orbits. 
This is unlike the case of idealised systems like $K$-flows where ({\it cf.} 
Anosov 1967) ${\Lambda}$ and ${\chi}$ are directly related. Why this is the 
case is not obvious. However, it seems reasonable to conjecture that it is the 
nontrivial character of the phase space, {\it i.e.,} the coexistence of both
regular and chaotic orbits and the existence of topological obstructions such
as the Arnold web, that renders phase mixing somewhat less efficient and 
substantially more complex than is the case for $K$-flows.

One thing that is evident is that the value of ${\Lambda}$ can depend on the
details. For example, ${\Lambda}$ can depend on the choice of phase space
coordinates that are being probed. Orbits can disperse faster in some 
directions than others, and this is reflected in the fact that convergence 
towards $f_{niv}$ can proceed at different rates for difference phase space
surfaces. Thus, {\it e.g.,} $f(x,y,t)$ and $f(x,z,t)$ can approach a
(near-)equilibrium at very different rates. Similarly, the value of ${\Lambda}$
can depend on the level of coarse-graining. Depending on the form of the 
potential and the choice of initial conditions, finer coarse-grainings, 
{\it i.e.,} a larger value of $k$ for the $k \times k$ grid, can yield either
larger or smaller values of ${\Lambda}$.

It is important to emphasise that the distribution $f_{niv}$ towards which the 
orbit ensemble evolves initially does {\it not} in general correspond to a
true microcanonical distribution, {\it i.e.,} a uniform population of the 
entire phase space region that is in principle accessible to the orbits; and,
as such it does not correspond to a strictly time-independent state. Indeed, if
the ensemble be evolved for much longer times one discovers oftentimes that
it will exhibit a slow secular evolution whereby it spreads further to probe
phase space regions which were avoided completely at earlier times. The 
original approach towards $f_{niv}$ typically proceeds on a time scale shorter 
than $10t_{D}$, with $t_{D}$ a characteristic dynamical time. These much 
slower variations can proceed on a time scale as long as ${\sim}{\;}1000t_{D}$
or even longer. What they reflect is a slow diffusion of chaotic orbits
through cantori in $M$-dimensional potentials with $M-1$ isolating integrals
or along the Arnold web in potentials with no more than $M-2$ isolating
integrals. These topological obstructions do not serve as absolute impediments
so as to prevent phase space transport. However, they {\it can} serve as 
partial barriers which significantly suppress the overall degree of phase 
space diffusion.

\subsection{Systems with a weak time-dependence}
Now consider how the idealised problem of chaotic mixing in a fixed 
time-independent potential can be impacted by weak time-dependent
irregularities, perturbations so weak that, over time scales of interest,
the values of the energy (and any other isolating integral associated with 
the unperturbed orbits) are very nearly conserved. The types of perturbations
which one might expect to be operative in real galaxies include:
\par\noindent
${\bullet}$ {\it periodic driving}, which can mimic the effects of companion
objects and/or systematic internal pulsations;
\par\noindent
${\bullet}$ {\it friction and white noise}, which can mimic discreteness
effects, including gravitational Rutherford scattering between individual
stars; and
\par\noindent
${\bullet}$ {\it coloured noise}, which can mimic the nearly random effects
of a high density cluster environment or incoherent internal pulsations.

These effects can be addressed computationally by generating numerical 
solutions to a {\it Langevin equation} of the form ({\it cf.} van Kampen 1981)
\begin{equation}
{d^{2}{\bf r}\over dt^{2}}=-{\nabla}V_{0}({\bf r})-
{\nabla}V_{1}({\bf r},{\omega}t) - {\eta}({\bf r},{\bf v}){\bf v}
+{\bf F}({\bf r},{\bf v},t).
\end{equation}
Here the first term on the right hand side represents the unperturbed
potential, whereas the second allows for time-dependent perturbations with
period ${\tau}=2{\pi}/{\omega}$. The third term is a dynamical friction and
the fourth corresponds to random kicks, which are described probabilistically.
Assuming in the usual fashion that the kicks correspond to Gaussian noise, 
everything about ${\bf F}$ is characterised by the first two moments, which are
assumed to satisfy
\begin{equation}
{\langle}F_{a}(t){\rangle}= 0 \qquad {\rm and} \qquad 
{\langle}F_{a}(t_{1})f_{b}(t_{2}){\rangle}=K_{ab}({\bf r},{\bf v},t_{1}-t_{2})
\qquad (a,b=x,y,z),
\end{equation}
with $K_{ab}$ the autocorrelation function. Picking $K_{ab}$ proportional
to a Dirac delta yield white noise, corresponding to instantaneous kicks with
a vanishing autocorrelation time. As a typical example of coloured noise, $K$ 
can be taken to sample the Ornstein-Uhlenbeck process, for which
\begin{equation}
K_{ab}{\;}{\propto}{\;}\exp(-|t_{1}-t_{2}|/t_{c}).
\end{equation}
It turns out that the relevant parameters for characterising the effects of
noisy perturbations are the autocorrelation time $t_{c}$ and diffusion
constant $D$, defined respectively by the relations 
\begin{equation}
t_{c}{\;}{\equiv}{\;}{\int dt\,tK(t)\over \int dt\,K(t)} \qquad {\rm and} 
\qquad D{\;}{\equiv}{\;}\int dt\,K(t)dt. 
\end{equation}
It is clear dimensionally that $D{\;}{\sim}{\;}F^{2}t_{c}$, where $F$
represents the characteristic amplitude of the random forces modeled by the
coloured noise.

Time-dependent perturbations of the form incorporated in this Langevin equation
can impact chaotic mixing (Pogorelov \& Kandrup 1999, Kandrup, Pogorelov, \&
Sideris 2000, Siopis \& Kandrup 2000) both (1) by accelerating the approach 
towards a
near-equilibrium in a single phase space region and (2) by facilitating
diffusion through cantori or along an Arnold web to achieve a true equilibrium.
The perturbations act via a {\it resonant coupling} between the
natural frequencies of the orbits and the natural frequencies of the 
perturbation. The largest effects arise when the perturbing frequencies
are comparable to $t_{D}^{-1}$, although there can also be couplings via 
harmonics and subharmonics. 

That periodic driving acts via a resonant coupling is hardly surprising.
That noise should also act via a resonant coupling is perhaps less obvious
but also true. The important point here is that the autocorrelation function
$K(t)$ determines the frequencies for which the perturbation has
power, since the spectral density is related simply to the Fourier transform
of $K$. White noise has a flat spectral density, with power at all frequencies,
and can thus couple to all frequencies. Coloured noise with a finite 
autocorrelation time $t_{c}$ cuts off at high frequencies so that, for
$t_{c}{\;}{\gg}{\;}t_{D}$, the noise has a comparatively minimal effect.

In the absence of time-dependent irregularities, phase mixing within a single
nearly disjoint phase space region is limited by Liouville's Theorem, which
constrains significantly the degree to which a swarm of orbits can `fuzz out'
on small scales, {\it e.g.} by guaranteeing that phase space trajectories do 
not cross. Even very weak perturbations can wiggle the orbits enough to allow 
such `fuzzing' to occur. 

Diffusion between nearly disjoint phase space regions is also facilitated 
by the fact that perturbations can wiggle the original orbits, thus helping 
them find appropriate avenues of escape through cantori or along the Arnold
web from one chaotic phase space region to another. Overall, the diffusion of
point masses through such phase space holes is well modeled as a Poisson
process, whereby orbits escape the original region at a near-constant rate.
In this sense, the physics is fundamentally similar to the problem of effusion 
of gas through a tiny hole in elementary statistical physics. The introduction 
of noise or periodic driving accelerates this process by continually wiggling
the orbits. 

It appears that the details of the perturbation are comparatively unimportant,
so that, {\it e.g.,} details that would be difficult (if not impossible) to
extract from observations are largely irrelevant. For example, 
different forms of noise, both additive and multiplicative, characterised by 
comparable amplitudes and autocorrelation times tend to have virtually
identical effects; and the presence or absence of dynamical friction, which
is a slowly varying systematic perturbation, tends to be nearly irrelevant. 
All that seems to matter are
%\par\noindent
%${\bullet}$ 
the amplitude of the perturbation, {\i.e.,} how hard the orbits
are kicked, and
%\par\noindent
%${\bullet}$ 
the autocorrelation time $t_{c}$, which sets the natural time scale
for the perturbation.

Quite generally, the overall efficacy of the perturbation as probed by (1)
the rate at which an ensemble approaches a near-equilibrium or (2) the rate at
which orbits diffuse through topological obstructions appears to scale
{\it logarithmically} with the diffusion constant $D$. For very large and very 
small autocorrelation times, the precise value of $t_{c}$ is nearly irrelevant:
for $t_{c}\to 0$ one recovers the results appropriate for white noise; for 
$t_{c}\to \infty$ the noise becomes almost completely irrelevant. For 
intermediate values of the autocorrelation time, the efficacy of the 
perturbation also scales {\it logarithmically} in $t_{c}$. 

It should be stressed that the perturbations do not act simply by making
the orbits more chaotic. Unless the perturbations are of comparatively large
amplitude, so large than energy is no longer approximately conserved, the
size of a typical Lyapunov exponent remains nearly unchanged. 

But how large must these perturbations be to have a signficant effect?
The answer here is that even very small perturbations can have a surprisingly
large effect. For example, white noise corresponding to a diffusion constant
$D{\;}{\sim}{\;}10^{-6}$ in natural units, and hence a relaxation time
$t_{R}{\;}{\sim}{\;}10^{6}t_{D}$, or coloured noise corresponding to kicks 
with $D{\;}{\sim}{\;}10^{-3}$ and time scale $t_{c}{\;}{\sim}{\;}
10t_{D}$ can increase the rate at which $f(t)$ approaches a near-invariant
$f_{niv}$ by a factor of three and the rate of diffusion through an Arnold 
web by an order of magnitude or even more. 

This behaviour is illustrated in
Figures 2 and 3 which, respectively, exhibit the effects 
of varying the
\begin{figure}[t]
\centering
\centerline{
        \epsfxsize=8cm
        \epsffile{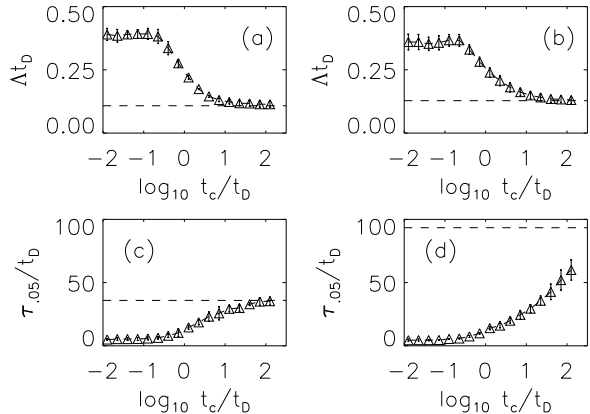}
           }
        \begin{minipage}{12cm}
        \end{minipage}
        \vskip -2.5in\hskip -0.0in
\caption{ %FIG. 1 
\small
(a) The convergence rate ${\Lambda}$ for an ensemble of chaotic
orbits in the lowest energy shell for the triaxial Dehnen potential with 
${\gamma}=1$, $c/a=1/2$ and $(a^{2}-b^{2})/(a^{2}-c^{2})=1/2$, allowing for
coloured noise with diffusion constant $D=2.5\times 10^{-4}$ and variable
autocorrelation time $t_{c}$. 
(b) ${\Lambda}$ for a different ensemble evolved in the same potential with
the same energy. (c) ${\tau}_{.05}$, the time required for $L^{2}$ convergence
towards $f_{niv}$ at the 5\% level on a $20 \times 20$ grid for the ensemble 
in (a). (d) The same for the ensemble in (b).
In each panel, the dashed line corresponds to the ensemble evolved without
any perturbations.} %\label{fig-2}
\vspace{0.2cm}
\end{figure}
\begin{figure}[t]
\centering
\centerline{
        \epsfxsize=8cm
        \epsffile{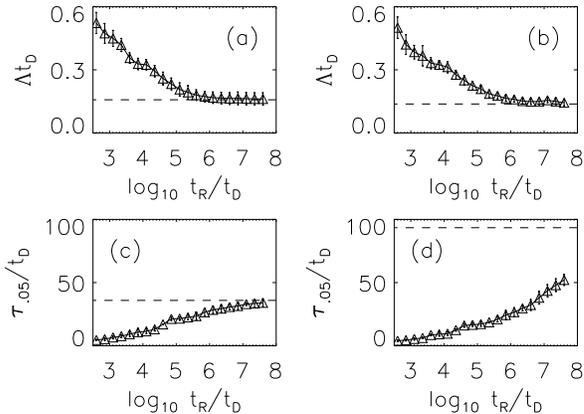}
           }
        \begin{minipage}{12cm}
        \end{minipage}
        \vskip -2.5in\hskip -0.0in
\caption{ %FIG. 1 
\small
(a) The convergence rate ${\Lambda}$ for an ensemble of chaotic
orbits in the lowest energy shell for the triaxial Dehnen potential with 
${\gamma}=1$, $c/a=1/2$ and $(a^{2}-b^{2})/(a^{2}-c^{2})=1/2$, allowing for
coloured noise with autocorrelation time $t_{c}=t_{D}$ and variable
diffusion constant $D$.
(b) ${\Lambda}$ for a different ensemble evolved in the same potential with
the same energy. (c) ${\tau}_{.05}$, the time required for $L^{2}$ convergence
towards $f_{niv}$ at the 5\% level on a $20 \times 20$ grid for the ensemble 
in (a). (d) The same for the ensemble in (b).
In each panel, the dashed line corresponds to the ensemble evolved without
any perturbations.} %\label{fig-3}
\vspace{-0.2cm}
\end{figure}
diffusion constant $D$ for fixed $t_{c}$  and the effects of varying $t_{c}$ 
for fixed $D$. The right and left panels exhibit results for two different 
orbit ensembles, in each case corresponding to a collection of orbits evolving 
in the lowest energy shell of the ${\gamma}=1$ triaxial analogues of the 
Dehnen potentials considered by Merritt \& Fridman (1996). In each figure, the 
top panels exhibit 
the initial convergence rate ${\Lambda}$ expressed in units of $t_{D}^{-1}$. 
The lower panels exhibit ${\tau}_{0.05}$, the time required for $L^{2}$ 
convergence towards $f_{niv}$ at the 5\% level. The ensemble in the left hand
panels corresponds seemingly to a `typical' ensemble of chaotic orbits in this
potential. The ensemble in the right hand panels is a somewhat less typical
ensemble for which the later time approach towards $f_{niv}$ was especially 
slow: even though the initial rates ${\Lambda}$ for the right and left 
ensembles are comparable initially, the right hand ensemble reaches a 
near-invariant $f_{niv}$ much more slowly. The obvious point then is that even 
very weak noise, corresponding to very large $t_{c}$ ($100t_{D}$ or longer) 
and/or very small $D$ ($2.5\times 10^{-4}$ or less) can 
dramatically accelerate the approach towards $f_{niv}$.
\section{Implications for Violent Relaxation}
Chaotic mixing has obvious implications for the rate at which various 
irregularities can disperse in a time-independent, or nearly time-independent,
potential. More important, however, is the fact that it could serve as an 
important ingredient in a satisfactory theory of violent relaxation. As 
described in Lynden-Bell (1967), violent relaxation relies on phase mixing 
which, {\it e.g.,} 
will cause an initially localised ensemble of points to disperse. The important
point, then, is that numerical experiments involving regular motions, in the 
spirit of Lynden-Bell's balls rolling in a pig-trough, yield comparatively
inefficient phase mixing, whereas chaotic motions in the same potential can 
yield a rapid, and effective, phase mixing ({\it cf}. Kandrup 1999).

It is, however, clear that realistic galactic potentials cannot be completely
chaotic, and one might anticipate that, unless chaotic orbits largely dominate
the flow, chaotic mixing need not prove all that important for violent 
relaxation. Nevertheless, there are several lines of reasoning that would 
suggest that chaotic mixing could prove of significant importance anyway: 
First of all, one knows that, generically, time-dependent potentials typically 
admit larger measures of chaotic orbits with sensitive dependence on initial 
conditions than do time-independent potentials. This is hardly surprising 
given the observation that they typically have one fewer global isolating 
integral. Time-dependent potentials also allow for the possibility of 
transitions between regularity and chaos. In particular, an orbit which is 
regular most of time can become chaotic for part of the time ({\it cf.}
Kandrup \& Drury 1998). As discussed above, it is clear that even very weak 
time-dependences can dramatically expedite phase mixing by facilitating both 
a `fuzzing' of orbits in a given, nearly disjoint, phase space region and a
diffusion between different phase space regions (Pogorelov \& Kandrup 1999, 
Kandrup \& Siopis 2000). For comparatively weak perturbations, one can 
think of the phase space as nearly unalterred and simply envision orbits being 
`assisted' in their phase space diffusion. However, larger perturbations can 
also blur, move, and (in some cases) even remove topological obstructions like 
the Arnold web, again facilitating accelerated phase space transport.

One final caveat should, however, be stressed: Chaotic mixing is likely to be
comparatively unimportant for structure formation in the early Universe, at 
least within the context of the standard Friedmann cosmologies. It would 
appear that the overall expansion of the Universe largely suppresses 
exponential chaos (Kandrup \& Drury 1998), so that chaotic mixing should be 
largely inoperative until a primordial irregularity has `pinched off' from the 
universal expansion.

\acknowledgments
I am pleased to ackowledge useful
interactions with my collaborators, Ilya Pogorelov, Ioannis Sideris, and,
especially, Christos Siopis and Elaine Mahon. This research was supported in 
part by NSF AST-0070809.

\vfill\eject
\end{document}